\documentclass[conference]{IEEEtran}
\IEEEoverridecommandlockouts
\usepackage{cite}
\usepackage{amsmath,amssymb,amsfonts}
\usepackage{algorithmic}
\usepackage{graphicx}
\usepackage{textcomp}
\usepackage{xcolor}
\usepackage{afterpage}
\usepackage{float}
\usepackage{booktabs}
\usepackage{makecell}
\usepackage[hidelinks]{hyperref}
\usepackage[nameinlink,noabbrev]{cleveref}
\def\BibTeX{{\rm B\kern-.05em{\sc i\kern-.025em b}\kern-.08em
    T\kern-.1667em\lower.7ex\hbox{E}\kern-.125emX}}
    
\begin{document}

\title{Spatial Audio Coding Through Relative Room Impulse Response Estimation}

\author{\IEEEauthorblockN{Nour Bouayed}
\IEEEauthorblockA{Orange Research, IRISA\\
Cesson-Sévigné, France \\
nourmeriem.bouayed@orange.com}
\and
\IEEEauthorblockN{Adrien Llave}
\IEEEauthorblockA{Orange Research\\
Cesson-Sévigné, France \\
adrien.llave@orange.com}
\and
\IEEEauthorblockN{Jérôme Daniel}
\IEEEauthorblockA{Orange Research\\
Lannion, France \\
jerome.daniel@orange.com}
\and
\IEEEauthorblockN{Pascal Scalart}
\IEEEauthorblockA{IRISA\\
Lannion, France \\
pascal.scalart@irisa.fr}
}

\maketitle

\begin{abstract}
Immersive virtual listening relies on spatial audio technologies such as Higher-Order Ambisonics (HOA), which represent sound scenes as multichannel signals. As the desired spatial resolution increases, so does the number of channels, making efficient compression essential for transmission over bandwidth-limited networks. Moreover, to facilitate deployment by network operators, the target bitrate for immersive audio coding should ideally remain close to the 25 kbps currently allocated to VoLTE audio services. State-of-the-art parametric codecs, such as the recently standardized Immersive Voice and Audio Services (IVAS) codec, achieve compression by transmitting spatial metadata together with a reduced number of transport channels. However, recent studies have shown that IVAS performance degrades on reverberant content, particularly at low bitrates, a limitation that suggests its inability to accurately model room acoustics. In this paper, we propose a novel HOA coding scheme based on the explicit and blind estimation of the Relative Spatial Room Impulse Response (ReSRIR), using a beamformed version of the HOA signal as a reference signal. By exploiting the structure and sparsity of the estimated ReSRIR, we derive an efficient parametric representation for immersive audio coding. Experimental evaluations show that the proposed method achieves higher compression than IVAS in the single-transport-channel regime, while maintaining comparable to slightly better quality.\footnote{Submitted to International Networked Immersive Audio 2026 (satellite Event of IEEE IS2 2026).}

\end{abstract}

\begin{IEEEkeywords}
Spatial audio compression, Parametric coding, Higher-Order Ambisonics.

\end{IEEEkeywords}

\section{Introduction}\label{sec:intro}
The human auditory system is built for three-dimensional listening: it can locate sounds, both in terms of direction and distance, and infer aspects of the surrounding environment geometry from the perceived reverberation. Spatial audio technologies aim at reproducing these perceptual cues through digital systems, enabling virtual listening experiences that feel close to natural hearing. Integrated into networking systems, these technologies gain even greater value, allowing users to share acoustic spaces with remote participants and to feel realistically present in a common environment such as a virtual meetings-room or a remote live event.

Among the available spatial audio representation formats, Higher-Order Ambisonics (HOA)~\cite{Daniel2001} offers a flexible and output system-agnostic representation of 3D sound scenes, making it particularly well suited for networked applications. Moreover, the scene is easy to manipulate; for example, rotation reduces to simple multiplication of the HOA components by a rotation matrix, enabling head-tracking in applications like multi-user virtual reality. However, within the HOA framework, achieving a good spatial resolution for localization requires a high channel count : at least $16$ channels (3\textsuperscript{rd}-order Ambisonics) are needed~\cite{Huisman2021}. This results in high data rates, which makes compression a necessary step towards practical deployment over bandwidth-constrained networks.

One common way of achieving multichannel signal compression is through parametric coding. It consists in extracting spatial parameters at lower time resolution (\emph{e.g.} 20~ms), compressed and transmitted along with a few decorrelated audio signals, called \textit{transport channels}. From them, the decoder reconstructs a perceptually convincing spatial sound scene. This paradigm is at the heart of the recently standardized Immersive Voice and Audio Services (IVAS) codec~\cite{Weckbecker2025a}, which builds on the Directional Audio Coding (DirAC)~\cite{Pulkki2017} and Spatial Audio Reconstruction (SPAR)~\cite{McGrath2019} methods. Despite this codec being considered at the state-of-the-art in spatial coding, a recent study~\cite{Llave2026a} comparing IVAS with a basic multi-mono approach demonstrated that IVAS performance tends to drop on reverberant audio for higher compression ratios. This suggests a limited ability to model the acoustic fingerprint of the room, commonly represented by Spatial Room Impulse Responses (SRIRs) which map the dry single-channel source signal to the reverberant multi-channel scene signal.

Parametric spatial codecs that use a single transport channel implicitly encode the Relative Spatial Room Impulse Response (ReSRIR; relative to that channel) in the transmitted spatial parameters, since it enables the reconstruction of the remaining channels from the transport channel. Examples include IVAS at 32~kbps and below, DirAC, and SpatialCodec~\cite{Xu2024}. Therefore, in the following, we focus on this single-transport-channel setting for interpretability and ease of comparison. It is also worth mentioning that bitrate should remain close to the currently available bitrate for VoLTE audio, \textit{i.e.}, 24~kbps, to be easily adopted by operators for immersive communications. This gives additional advantage to lower bitrate versions of IVAS (32~kbps, 64~kbps) over higher ones (256~kbps, 512~kbps).


In this single-transport-channel regime, IVAS spatial parameters are predictive coefficients that relate each HOA component to the transport channel in the time-frequency domain, 
\textit{i.e.}, an estimate of the Relative Transfer Function (RTF), the frequency-domain counterpart of the time-domain ReSRIR.

In the current design, only the real part of these coefficients is retained~\cite{etsi_wki_78490}, which translates into a symmetric (and thus anti-causal) ReSRIR in the time domain. In addition, reverberation in IVAS is largely synthesized using spatial decorrelation filters. A mismatch between these filters and the true room response can degrade quality, especially because such filters are better suited to late, noise-like reverberation than to accurately reproducing early reflections. Yet, the latter strongly influence perceived quality and spatial impression~\cite{Greco2025, Shlomo2021}.

It is worth mentioning that some recently proposed neural-network-based spatial audio codecs (also referred to as neural audio codecs) adopt a similar parametric coding paradigm~\cite{Kleijn2023}. For example, SpatialCodec~\cite{Xu2024} learns to estimate and encode spatial filters called Complex Ratio Filters (CRFs), which represent a convolutive RTF that maps a reference channel to the other channels in the time-frequency domain. However, SpatialCodec was proposed for signals captured by a linear (non-coincident) microphone array, with the goal of preserving spatial information for beamforming after decoding. Adapting such a method to Ambisonics is not straightforward, and their objective of enabling post-decoding beamforming does not fully align with the goal of preserving the perceived quality of the sound scene. Other neural spatial codecs include multichannel extensions of state-of-the-art single-channel neural audio codecs, notably a multichannel version of DAC for 3\textsuperscript{rd}-order ambisonics~\cite{Hirvonen2024} (trained on ambient content) and FOA Tokenizer, a multichannel version of WavTokenizer for 1\textsuperscript{st}-order ambisonics~\cite{Sudarsanam2025}. Unlike the parametric approaches discussed above, these models do not explicitly separate spatial filters from excitation (transport) signals in their coding architecture. We do not consider these approaches here because of the unavailability of open-source code and model weights for reproducible evaluation.

In this paper, we propose a new coding scheme for HOA signals based on the explicit and blind estimation of the Relative Spatial Room Impulse Response (ReSRIR). Following~\cite{Kitic2024a}, we use a beamformed version of the HOA signal as a reference for this estimation. We then exploit the structure and sparsity of the resulting ReSRIR to obtain an efficient representation that enables higher compression while achieving quality comparable to IVAS in reverberant conditions. The remainder of this paper is organized as follows. Section~\ref{sec:method} details the proposed approach, including the estimation and the parameterization of the ReSRIR. The experimental evaluation against existing methods is presented in Section~\ref{sec:experiments}, and Section~\ref{sec:conclusion} concludes with our findings and outlines future work.

\section{Method}\label{sec:method}

\begin{figure}[t]
    \centering
    \includegraphics[width=0.8\linewidth]{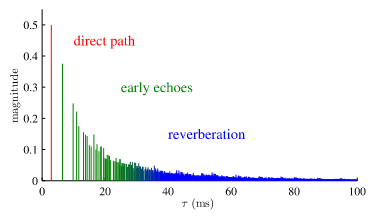}
    \caption{Schematic illustration of the shape of an RIR (from \cite{Gannot2017}).}
    \label{fig:rir}
\end{figure}

\begin{figure*}[t]
  \centering
  \includegraphics[width=\textwidth]{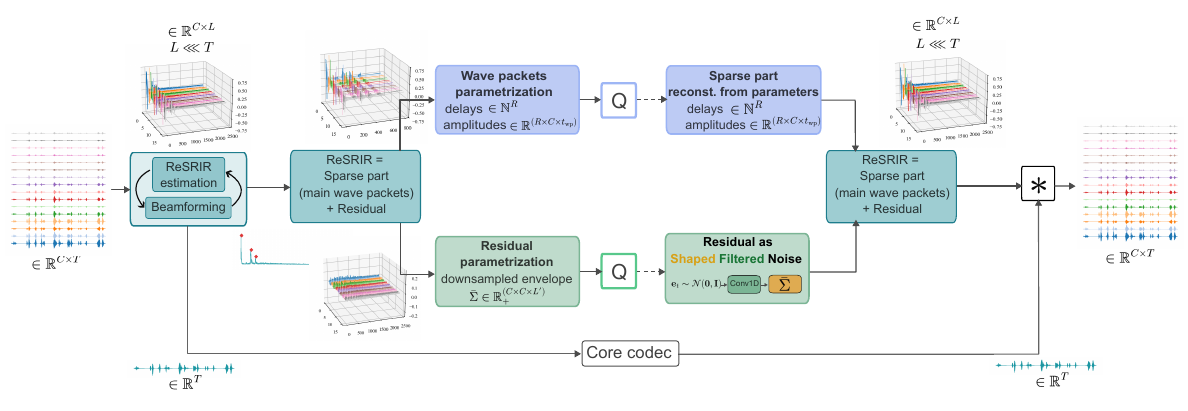}
  \caption{Block diagram of the proposed coding method.}
  \label{fig:method-diagram}
\end{figure*}

\subsection{Signal model}
Let $t$ denote the time index, and $\mathbf{b}(t) \in \mathbb{R}^{C}$ the multichannel signal of spherical harmonic (SH) expansion coefficients up to order $O$ (with $C=(O+1)^2$ in the 3D case), also referred to as the HOA components of the sound scene.

Considering the sound scene is due to a single static source $s(t)$ in a noisy reverberant environment, the HOA signal $\mathbf{b}(t)$ can be formulated in the time domain as follows:
\begin{equation}
\mathbf{b}(t) = \mathbf{x}(t) + \mathbf{n}(t),
\end{equation}
where $\mathbf{x}(t)$ is defined as the source \textbf{spatial image}~\cite{Gannot2017} and $\mathbf{n}(t) \in \mathbb{R}^{C}$ is the ambient noise.

\subsubsection*{Terminology (RIR, SRIR, ReSRIR)}\label{subsubsec:Terminology} A Room Impulse Response (\textbf{RIR}) characterizes sound propagation from a source to a microphone, such that convolving it with the source signal yields the reverberant microphone signal. As sketched in Figure~\ref{fig:rir}, an RIR typically comprises a direct-path peak, a set of early reflections, and a late-reverberation tail.
In multichannel spatial audio, the RIR corresponding to each HOA component can be stacked to form a Spatial RIR (\textbf{SRIR}).
In the HOA framework,  components can be viewed as outputs of virtual coincident microphones with spherical-harmonic directivities. Therefore, a given reflection arrives simultaneously to all of them and appears as a time-aligned peak across the SRIR components.
In practice, the clean source is not available, so it is more convenient to consider the Relative SRIR (\textbf{ReSRIR}) to a \emph{pseudo-source}, \emph{e.g.}, a dereverberated estimation of the clean source.

In practice, SRIRs are represented using Finite Impulse Responses (FIR) filters of length $L$ taps. Then, the spatial image can be modeled as the convolution of the clean source signal $s(t)\in \mathbb{R}$ with the SRIR $\mathbf{a}(t)\in \mathbb{R}^C$:
\begin{align}
\mathbf{x}(t) &= (\mathbf{a} * s)(t), \\
              &= \sum_{\tau=0}^{L-1}\mathbf{a}(\tau)  s(t-\tau). \label{eq:model_time}
\end{align}
Separating $\mathbf{a}(t)$ into an early directional component, gathering the direct path and early reflections, and a late diffuse component for the remaining dense reverberation, it can be expressed as:
\begin{equation}
    \mathbf{a}(t) = \mathbf{a}_\text{early}(t)+\mathbf{a}_\text{late}(t),
\end{equation}
with
\begin{equation}
    \mathbf{a}_\text{early}(t) =
\sum_{n=0}^{N-1} \nu_n(t)\,\mathbf{y}_n(\Omega_n)\,\delta(t-\tau_n),
\label{eq:SRIR_early}
\end{equation}
Where $\nu_n(t)\in(0,1)$ is the attenuation factor, $\tau_n\in\mathbb{R}_+$ is the time-of-arrival (ToA), and $\mathbf{y}_n(\Omega_n)\in\mathbb{R}^{C}$ is the real-valued SH encoding vector associated to the $n$\textsuperscript{th} acoustic plane wave impinging from direction $\Omega_n$. The Dirac delta $\delta(t-\tau_n)$ models each early reflection as an impulse located at delay $\tau_n$. The plane-wave expansion is truncated to $N$ wavefronts, while the residual late reverberation is captured by the additive term $\mathbf{a}_\text{late}(t)$.

In the conventional Short-time Fourier Transform (STFT) domain, let $\ell$ denote the frame index and $f\in\{0,\dots,F-1\}$ the frequency-bin index (with $F$ frequency bins). Time-domain convolution with a time-invariant SRIR translates into inter-frame and inter-band convolution with an Acoustic Transfer Function (ATF) \cite{Gannot2017}.
However, provided that the frame length is sufficiently large compared to the RIRs length, the Multiplicative Transfer Function (MTF) approximation can be applied and the time-domain model in \eqref{eq:model_time} simply becomes in the STFT domain:
\begin{equation}
\mathbf{x}(\ell,f)= \mathbf{a}(f) \cdot s(\ell,f), \label{eq:MTF}
\end{equation}
with $\mathbf{x}(\ell,f)\in\mathbb{C}^C$, $\mathbf{a}(f)\in\mathbb{C}^C$ and $\mathbf{s}(\ell,f)\in\mathbb{C}$.
In practice, when source signal is not accessible, the multichannel HOA signal is modeled using an RTF applied to a known reference channel, denoted $\mathbf{a_\text{r}}(f)$ and $x_r(\ell,f)$ respectively:
\begin{equation}
\mathbf{x}(\ell,f)= \mathbf{a_\text{r}}(f) \cdot x_r(\ell,f)\text{.}
\label{eq:RTF}
\end{equation}
Reference channel can be one of the HOA signal components, or any other signal correlated with the source, for example a beamformed version of the HOA signal.

\subsection{Overview of the proposed method}
Based on the signal model in Equation \eqref{eq:RTF}, the proposed coding method follows a three-stage pipeline, as illustrated in Figure \ref{fig:method-diagram}. First, we estimate a spatial room impulse response relative (ReSRIR) to a beamformed version of the HOA signal, respectively $\mathbf{a_\text{r}}$ and $x_r$. Second, the estimated ReSRIR is decomposed into a \textit{sparse} part, which contains the dominant propagation paths, and a \textit{residual}, which contains the remaining diffuse energy. This decomposition is further parametrized, and the resulting parameters are quantized as a final encoding step. At the decoder side, the ReSRIR is re-synthesized from its parameters and applied through convolution to the reference signal $x_r$, which has been independently coded using a core monochannel audio codec.

\subsection{ReSRIR estimation using GTVV}
The first step is to estimate the ReSRIR in the time domain, or equivalently the RTF in the frequency domain. Assuming the MTF approximation holds, a common HOA RTF estimator on the $\ell$\textsuperscript{th}-frame and $f$\textsuperscript{th}-frequency-bin is the Velocity Vector $\mathbf{\hat{a}_\text{r}}(\ell,f)$ defined as : 
\begin{equation}
\mathbf{\hat{a}_\text{r}}(\ell,f)=\frac{\mathbf{x}(\ell,f)}{x_0(\ell,f)},
\end{equation}
where $x_0(\ell,f)$ denotes the first HOA channel of the spatial source image (\textit{i.e.} the omnidirectional component), which is used as a reference signal for the RTF. Replacing this reference with the output of a beamformer $\mathbf{w}(k)\in \mathbb{C}^C$ steered toward the source direction, one obtains the Generalized Frequency-domain Velocity Vector (GFVV), as defined in~\cite{Kitic2024a}:
\begin{equation}
\mathbf{\hat{a}_\text{r}}(\ell,f)=\frac{\mathbf{x}(\ell,f)}{\mathbf{w}(k)^H\mathbf{x}(\ell,f)}.\label{eq:GFVV}
\end{equation}
Its time-domain counterpart, the Generalized Time-Domain Velocity Vector (GTVV), which is the ReSRIR, is sparse and quasi-causal when the reference signal is dominated by direct propagation, \textit{i.e.}, when the beamformer is sufficiently selective toward the direct sound and it attenuates reverberation~\cite{Kitic2024a}.

The ReSRIR is estimated using the "self-steering" algorithm from \cite{Kitic2024a}, which is based on the RTF estimation method that leverages speech non-stationarity, defined in~\cite{Jarrett2017}. The key idea is that it is an iterative procedure that jointly estimates the ReSRIR and the source DoA. Starting with the omnidirectional HOA component as a reference signal (that is, $\mathbf{w} = [1,0,\ldots,0]^T$), the algorithm uses the resulting ReSRIR at zero-delay ($\mathbf{\hat{a}_\text{r}}(t=0)$) to obtain an initial DoA estimate. This estimate is then used to design a more selective beamformer for the next iteration, progressively refining both the DoA and the ReSRIR estimate until convergence. In our experiments, five iterations were sufficient for convergence.

Once the RTF (GFVV) has been estimated, we compute its time-domain representation, the ReSRIR (GTVV). To mitigate circular-convolution artifacts, we zero-pad the signals prior to the Fourier transform and truncate anti-causal part of the result after the inverse Fourier transform.

\begin{figure}[t]
    \centering
    \includegraphics[width=\linewidth]{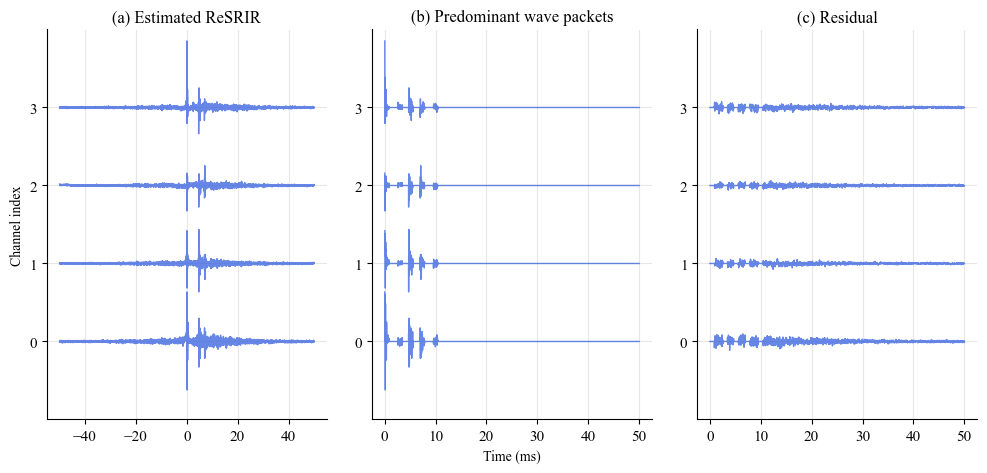}
    \caption{Example of a ReSRIR estimated on a real $3^\text{rd}$-order HOA speech signal (leftmost plot (a)). Only first four channels are displayed for clarity, and only causal part is kept and shown after decomposition into sparse component (middle plot (b)) and residual component (rightmost plot (c)).}
    \label{fig:GTVV_example}
\end{figure}

\begin{figure}[t]
    \centering
    \includegraphics[width=0.9\linewidth]{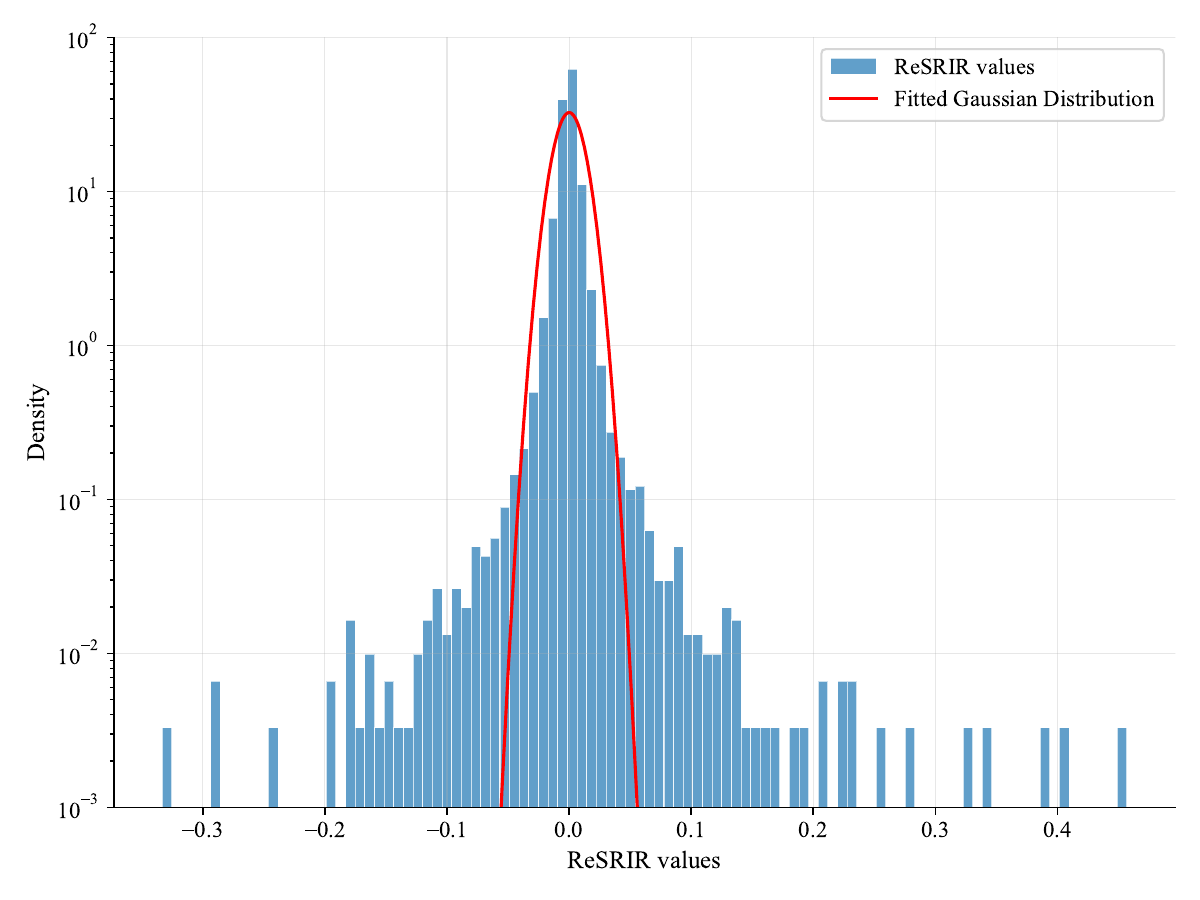}
    \caption{Histogram (log-scaled) of magnitudes taken by an estimated ReSRIR and the fitted Gaussian distribution. The mismatch between them depicts the excess Kurtosis and therefore the sparse nature of the ReSRIR.}
    \label{fig:GTVV_sparsity}
\end{figure}

An example of the ReSRIR returned by the estimation step is given in Figure~\ref{fig:GTVV_example}.
As can be seen, the ReSRIR is sparse and quasi-causal.
The histogram of its values shown in Figure~\ref{fig:GTVV_sparsity} further supports its sparse nature.
To compress it, we choose to decompose it into a dominant sparse signal and a residual signal.
 
\subsection{ReSRIR decomposition and parametrization}
The sparse part of the ReSRIR contains its main peaks, which share the same time positions (delays) across HOA components, as explained in Subsection~\ref{subsubsec:Terminology}. Therefore, to parametrize this part, we extract information about its $R$ main peaks: namely, their delays and their multichannel magnitudes.
To do so, we rely on the delay-magnitude curve, denoted $\zeta_v$, defined as follows~\cite{Kitic2024a}:
\begin{equation}
    \zeta_\mathbf{\hat{a}_\text{r}}(t) = \lVert \mathbf{\hat{a}_\text{r}}(t)\rVert_2.
\end{equation}
Peak delays are defined as time positions where $\zeta_v$ is locally maximized, and peak magnitudes are simply ReSRIR values at those delays.
Because of the surface absorption, those peaks act as short impulse responses.
Therefore, we represent each one by a wave packet spanning $t_\text{wp}$ samples, instead of relying on a single instantaneous sample.

Hence, we obtain two tensors to represent the sparse part: a tensor of multichannel wave-packet $\mathbf{P} \in \mathbb{R}^{R \times C\times t_\text{wp}}$, and a vector of corresponding delays $\mathbf{d} \in \mathbb{N}^{R}$.
In this work, we experiment with two compression strategies :
\begin{enumerate}
    \item \textbf{Projection/expansion with linear layers}. Each wave packet at position $r$,  $\textbf{P}^{(r)} \in \mathbb{R}^{C\times t_\text{wp}}$ is projected onto a smaller $d$-dimensional space then expanded back using two sequential linear layers, yielding $\hat{\mathbf{P}}^{(r)}\in \mathbb{R}^{C\times t_\text{wp}}$. Combined with the vector quantization (VQ) step in Section~\ref{subsec:quantization}, this is equivalent to the improved-VQ layer introduced in \cite{Yu2022}. 
    \item \textbf{Plane wave decomposition}. Each wave packet is approximated by a single vector (Dirac impulse) given by the spherical harmonic encoding vector that is most correlated with its pseudo-intensity vector, based on the model in Equation \eqref{eq:SRIR_early}. Although this technique preserves the DoA of reflections that are synthesized by the ReSRIR, it is no longer able to model the spectral modifications between the direct path and the early reflections, as well as between the reference signal and the HOA components.
\end{enumerate}

The residual multichannel signal $\mathbf{r}\in\mathbb{R}^{C\times L}$ is obtained by subtracting the sparse component from the initial estimated ReSRIR. We parameterize it via short-time covariance matrices $\bar{\Sigma} \in \mathbb{R}^{C\times C\times L'}$, where $L'$ denotes the number of short-time segments. In practice, the estimated covariances are close to diagonal; therefore, we use the per-channel delay-dependant variances, denoted $\sigma_{c,k}^2$ ($c=1,\ldots,C$, $k=1,\ldots,L'$), \emph{i.e.}, a downsampled multichannel envelope $\sigma \in \mathbb{R}^{C\times L'}$ with $L'\ll L$.

At the decoder, we synthesize the residual using filtered noise shaping~\cite{Steinmetz2021}. We first generate multivariate Gaussian white noise and pass it through learned 1D convolutional filters trained to minimize a spectral reconstruction loss on residuals. We then apply the estimated shaping parameters (\emph{i.e.}, the envelope $\sigma$ upsampled via linear interpolation) to shape the filtered noise over time, thereby reconstructing an approximation of the residual $\mathbf{\hat{r}}\in\mathbb{R}^{C\times L}$. Inspecting the residual spectra suggests that the learned filters must act as high-pass filters for higher-order components. This may be linked to how ambisonics encoding from spherical microphone array naturally attenuates low frequencies in those components~\cite{moreau_3d_2006}.

\subsection{Parameters quantization}\label{subsec:quantization}
Following the ReSRIR parameterization, we compress the extracted parameters using learned quantizers, optionally preceded and followed by learned projection/expansion layers:
\begin{itemize}
\item \textbf{Delay differences:} we apply scalar quantization to differential delays after sorting the delays in ascending order;
\item \textbf{Wave-packet amplitudes:} we apply vector quantization either to the flattened wave packets (optionally after dimensionality reduction via a learned projection) or to the equivalent spherical-harmonics encoding vector, depending on the chosen parameterization;
\item \textbf{Residual envelope:} we apply vector quantization to the $C$-dimensional vector of residual-envelope values at each downsampled time index.
\end{itemize}

\subsection{Hyperparameters and training}
\subsubsection{Architecture}
We implement the proposed model using the following configuration: ReSRIR estimation is performed on frames of length $2048$ samples (\emph{i.e.}, $42\,\mathrm{ms}$). We extract $R=3$ wave packets from each estimated ReSRIR and set the wave-packet support to $t_\text{wp}=42$ samples. For wave-packet coding, we learn a linear projection that maps each flattened multichannel wave packet (dimension $C . t_\text{wp}=16\times 42=672$) to a $128$-dimensional latent space prior to quantization. The residual envelope is downsampled to $L'=7$ time indices. We learn vector quantizers with resolutions of $5$, $6$, and $11$ bits for the differential delays, residual-envelope vectors, and projected wave-packet latents, respectively. This yields a total of $11 + 2\times(5+11) + 7\times 6 = 85$ bits per frame, \emph{i.e.}, approximately 2~kbps for spatial-parameter coding.
\subsubsection{Training data}
For our experiments, we synthesize $3^\text{rd}$-order Ambisonics signals sampled at 48~kHz using real measured SRIRs. We combine three public databases of HOA SRIRs: Aalto~\cite{McKenzie2022}, IKSLabIR~\cite{Chatzimoustafa2025}, MOTUS~\cite{Gotz2021}. These SRIRs are truncated to 40~ms-long responses to satisfy the MTF approximation in Equation \eqref{eq:MTF}. Then, the SRIRs are convolved on-the-fly with random $3\text{s}$-excerpts of speech from DNS-Challenge dataset \cite{Reddy2020}. We further apply data augmentation by randomly rotating the scenes in 3D.
\subsubsection{Training method}
We train the proposed model in two stages. In the first stage, the quantization layers are bypassed and the remaining modules are optimized using the following reconstruction losses:
\begin{itemize}
\item \textbf{Wave packets projection/expansion loss.}
We use a Mean Square Error (MSE) loss between original and reconstructed wave packets:
\begin{equation}
\mathcal{L}_\mathrm{wp}(\mathbf{P},\hat{\mathbf{P}}) = \sum_{r, c, t'} \left(P^{(r)}_{c,t'}-\hat{P}^{(r)}_{c,t'}\right)^2.
\end{equation}

\item \textbf{Residual spectral reconstruction loss.} We use an MSE loss between spectral magnitudes of original and decoded residuals (through filtering):
\begin{equation}
    \mathcal{L}_{\text{residual-filters}}(\mathbf{r}, \mathbf{\hat{r}}) = \sum_{c=1}^C\rVert |\text{FFT}(r_c)| - |\text{FFT}(\hat{r}_c)| \lVert^2_2.
\end{equation}
\end{itemize}

In the second stage, we freeze all previously trained modules and train only the quantization layers to learn quantization codebooks. For this purpose, we optimize, for each quantizer, an MSE loss between the input vectors and their quantized version.

The codec is trained using Adam optimizer with an initial learning rate of $5\times10^{-3}$ and a batch size of $32$. We perform $700$ optimization steps during the first training phase, and $1500$ steps during the second training phase.

\section{Evaluation}\label{sec:experiments}

\subsection{Evaluation methods}
\subsubsection{Data}
We follow the same procedure as for training data generation, using SRIRs from two other public datasets : RSoANU~\cite{Chesworth2024} and 50-Lebedev \cite{Turner2023}. Speech signals are also drawn from DNS-Challenge dataset, from a different portion than the training set, in $10\text{s}$-long excerpts. We build test sets of $50$ signals for the evaluation of models.

\subsubsection{Baselines}

We compare the proposed method against IVAS~\cite{noauthor_codec_2024} at 32~kbps, because at this bitrate IVAS operates in single-transport-channel mode, enabling a fair comparison with our approach. To isolate the effect of spatial parameters coding, which is the focus of our study, we use the same core codec (EVS) at the same bitrate (24.4~kbps) as IVAS on the reference signal. We also report results for IVAS at 64~kbps for an additional reference point.

Table~\ref{tab:bitrate-allocation} summarizes the bitrate allocation for each of the aforementioned methods and our proposed approach.

\begin{table}[b]
    \caption{Bitrate allocation across transport channels and spatial parameters for each model.}
    \label{tab:bitrate-allocation}
    \begin{tabular}{l c c c}
        \toprule
         & \makecell{Proposed \\@27kbps} & \makecell{IVAS \\@32kbps} & \makecell{IVAS \\@64kbps} \\
        \midrule
        1\textsuperscript{st} transport channel & 24 & 24 & 38 \\
        2\textsuperscript{nd} transport channel & -- & -- & 16 \\
        Spatial parameters & 2 & 7 & 9 \\
        Additional budget for headers & 1 & 1 & 1 \\
        \midrule
        Total (kbps) & 27 & 32 & 64 \\
        \bottomrule
    \end{tabular}
\end{table}

\subsubsection{Criterion}
To evaluate the performance of tested methods, we rely mainly on QASTANet, a deep learning-based metric for evaluating overall spatial audio quality~\cite{Llave2025, code_qastanet_2025}. 

\subsection{Results}

\begin{figure*}[htb]
    \centering
    \includegraphics[width=\linewidth]{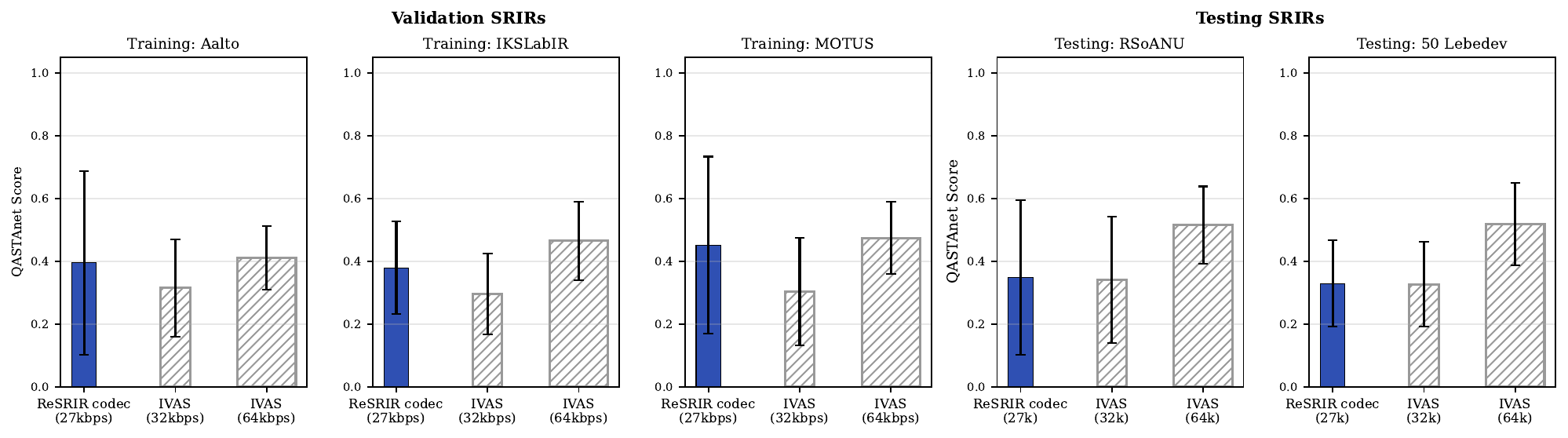}
    \caption{Mean QASTANet spatial audio quality scores on tested models (IVAS and proposed model) and on various testsets (each containing 50 samples). The vertical lines denote the standard deviation. Bar width is proportional to bitrate.}
    \label{fig:comparison-ivas}
\end{figure*}

\begin{figure*}[ht]
    \centering
    \includegraphics[width=\linewidth]{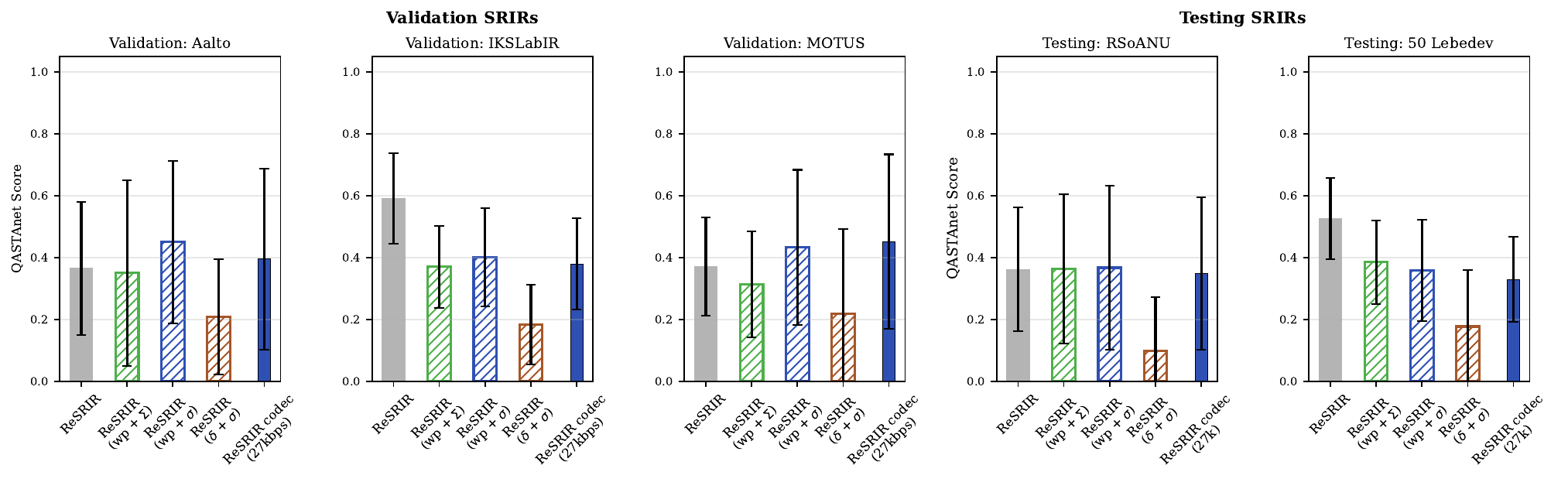}
    \caption{QASTANet scores averaged across items for various parametrization strategy, see text. The vertical lines denote the standard deviation.}
    \label{fig:ablation}
\end{figure*}

\subsubsection{Comparison to baselines}

Figure~\ref{fig:comparison-ivas} reports the mean QASTANet spatial audio quality scores for the proposed method and the considered baselines across the evaluated testsets. Overall, the proposed method achieves comparable scores to IVAS 32~kbps at a lower bitrate, on the two unseen SRIR datasets during the training RSoANU and 50-Lebedev. On validation data, which was generated based on the same SRIRs sources as the training data but using data augmentations such as random gains application and rotations, as well as speech sample and SRIR pairing, it even surpasses it with an improvement of the QASTANet score of 0.1, approaching IVAS 64~kbps scores. Reconstruction examples from our codec are provided on the demo page.\footnote{\url{https://nour-m-bouayed.github.io/Demo-IWNIA-2026/}}
\subsubsection{Ablation study}
To assess the contribution of our parameterization design choices, we perform an ablation study that compares the following conditions (with quantization step being bypassed):
\begin{itemize}
    \item \textbf{ReSRIR}: direct reconstruction from the estimated ReSRIR;
    \item \textbf{ReSRIR} ($\text{wp}$ + $\bar{\Sigma}$): residual modeled with full covariance matrices, and sparse component modeled with projected/expanded wave packets;
    \item \textbf{ReSRIR} ($\text{wp}$ + $\sigma$): residual modeled with diagonal covariance matrices (envelopes), and sparse component modeled with projected/expanded wave packets;
    \item \textbf{ReSRIR} ($\delta$ + $\sigma$): residual modeled with diagonal covariance matrices (envelopes), and sparse component modeled with equivalent SH encoding vectors as Dirac impulses.
\end{itemize}
The results are summarized in Figure~\ref{fig:ablation}.
Overall, the chosen parametrization configuration \textbf{ReSRIR (wp + $\sigma$)} is superior, yielding performance results close to reconstruction from non-approximated estimated ReSRIR.
Surprisingly, using full covariance matrices for the residual (\textbf{ReSRIR (wp + $\bar\Sigma$)}) does not improve performance and can even degrade it compared to the simpler diagonal-covariance (envelope) model (\textbf{ReSRIR (wp + $\sigma$)}). A plausible explanation is that a broadband covariance description is too coarse to capture the frequency-dependent statistics of the ReSRIR. As coding a more faithful multiband covariance model would introduce a parameter overhead, we leave this exploration for future work. Finally, replacing the wave-packet sparse component with Dirac-like impulses (\textbf{ReSRIR ($\delta$ + $\sigma$)}) significantly degrades performance. This was also observed during informal listening sessions with an increase in spectral distortions.
Indeed, this approximation of the wave-packet as a Dirac-impulse is unable to model the spectral mismatch between the direct path and the early reflections due to wall absorption.

\section{Conclusion}\label{sec:conclusion}
In this work, we presented a new method for efficiently coding higher-order ambisonics audio, based on blind estimation of the relative room impulse response and its parameterization. We evaluated it under controlled conditions against the IVAS codec. The results indicate that the proposed approach achieves comparable to slightly better performance at lower bitrates. Overall, the study supports the view that careful modeling and coding of spatial parameters can improve spatial rendering quality under bitrate constraints. Our investigations were limited to single-source scenes; extending the framework to multiple sources would increase the number of spatial parameters and transport channels, but in many practical scenarios sources rarely overlap strongly, and the slow temporal evolution of SRIRs could be exploited to reduce the parameter bitrate. Future work will extend the evaluation to more diverse acoustic conditions and content, and will include comprehensive subjective listening tests.

\bibliographystyle{IEEEtran}
\bibliography{references}

\end{document}